\documentclass[
    ,final            
  ]
  {aipproc}

\layoutstyle{6x9}

\begin{document}

\title{Metal-poor Galaxies in the Local Universe}

\classification{98.35.Gi, 98.56.Wm, 98.56.Si, 98.62.Ai, 98.62.Bj, 98.62.Lv 
                }
\keywords      {Galactic halo, 
                Dwarf galaxies (elliptical, irregular, and spheroidal),
                Magellanic Clouds and other irregular galaxies,
                Origin, formation, evolution, age, and star formation,
                Chemical composition and chemical evolution,
                Stellar content and populations; radii; morphology and overall structure
                }

\author{Eva K.\ Grebel}{
  address={Astronomisches Rechen-Institut,
  Zentrum f\"ur Astronomie der Universit\"at Heidelberg,
  M\"onchhofstr.\ 12--14, 69120 Heidelberg, Germany}
}

\begin{abstract}
A galaxy's mean metallicity is usually closely correlated with its
luminosity and mass.  Consequently the most metal-poor galaxies in the
local universe are dwarf galaxies.  Blue compact dwarfs and tidal
dwarfs tend to deviate from the metallicity-luminosity relation by
being too metal-poor or too metal-rich for their luminosity,
respectively.  A less pronounced offset separates dwarf spheroidal
(dSph) and dwarf irregular galaxies, making the former too metal-rich
for their luminosity, which indicates different formation conditions
for these two types of dwarfs.  While environment (photo-evaporation
through local re-ionization by massive galaxies, tidal and ram
pressure stripping) govern the observed morphology-distance relation,
intrinsic properties (in particular total mass) play a decisive role
in dwarf galaxy evolution with respect to the time and duration of
star formation and the amount of enrichment.  The metallicity
distribution functions of nearby dwarfs can be understood taking
pre-enrichment, gas infall, and winds into account.  Many dwarfs show
evidence for inhomogeneous, localized enrichment.  Ultra-faint dSphs,
which may have formed their metal-poor stars at high redshift via
H$_2$ cooling, show an overabundance of metal-deficient stars as
compared to the (inner) Galactic halo, but may, along with classical
dSphs, have contributed significantly to the build-up of the outer
halo.  The abundance ratios measured in the irregular Large Magellanic
Cloud are consistent with the postulated early accretion of irregulars
to form the inner Galactic halo. 
\end{abstract}

\maketitle


\section{1. Introduction}

Most galaxies in the local universe follow a metallicity-luminosity
(or metallicity-mass) relation in the sense that more luminous
galaxies also tend to be more metal-rich.  Galaxies in the
high-metallicity, high-mass regime with current stellar masses greater
than $10^{10}$~M$_{\odot}$ show very little increase in metallicity
towards higher masses though \cite{Panter2008}, while there is a
pronounced decline in metallicity towards lower luminosities (or
stellar masses) and an increased scatter in the relation
\cite{Panter2008}.

Taking three parameters into account, namely the stellar mass, the
gas-phase metallicity, and the star formation rate, galaxies up to a
redshift of $z \sim 2.5$ follow a fairly tight ``fundamental
metallicity relation'' in this three-dimensional space
\cite{Mannucci2010}.  At higher redshifts, galaxies show lower
metallicities leading to increasing deviations from the relation.
Here dilution by infall of pristine gas becomes a dominant factor
despite active star formation as long as the dynamical time scales
remain shorter than the chemical enrichment time scales
\cite{Mannucci2010}.  At lower redshift outflows of enriched material
become more important and a balance between infall and outflows
develops \cite{Mannucci2010}, whereas interacting and merging galaxies
deviate from the stellar mass-metallicity trend
\cite{Michel-Dansac2008}.  

In this review, we consider metal-poor galaxies in the {\em local}
Universe, i.e., galaxies at the low-luminosity, low-mass end of the
metallicity-luminosity relation --- thus, dwarf galaxies.  Exactly
where to draw the line between ``dwarf'' and ``giant'' galaxies is
more a matter of definition than of physical difference.  Often a
simple luminosity criterion is used (e.g., a luminosity $\leq 0.1\,
L_{\star}$ or $M_V \leq -18$ mag; see, e.g., \cite{Grebel2000}), which
also excludes the massive galaxies along the classical Hubble
sequence.  

Dwarf galaxies cover a variety of morphological types, luminosities,
present-day star formation activity, gas content, as well as
environments \cite{Grebel2001}.  Late-type, gas-rich, star-forming
dwarfs (such as dwarf spirals, dwarf irregulars, or blue compact
dwarfs) are usually found in low-density environments like the outer
regions of galaxy clusters, groups, or in isolation in the field.
Early-type, gas-deficient, quiescent dwarfs (such as dwarf
ellipticals, dwarf spheroidals, or ultra-compact dwarfs) are primarily
found in high-density environments, e.g., in the vicinity of massive
galaxies in groups and clusters.  This morphology-distance or gas
fraction-distance relation \cite{Grebel2003, Grebel2005} of the dwarf
galaxies perpetuates the morphology-density relation observed for
giant early-type and late-type galaxies (e.g., \cite{Oemler1974,
Dressler1980}) and suggests that environmental effects may play an
important role in shaping the evolution of dwarf galaxies (e.g.,
\cite{vandenBergh1994, Grebel2003}).

The best-studied dwarfs are the dwarf galaxies of the Local Group,
which are sufficiently close be resolved into individual stars and to
even permit us detailed studies of their stellar content down to their
oldest main-sequence turn-offs (see \cite{Monelli2010a, Monelli2010b,
Hidalgo2011} for recent examples).  All nearby dwarf galaxies studied
in detail so far, regardless of their morphological type, contain very
old stars with ages $> 10$ Gyr \cite{Grebel2004, Glatt2008a,
DaCosta2010}.  Though the fractions of these old populations vary and
more massive dwarf galaxies usually show more extended episodes of
star formation, old populations (i.e., early star formation) appear to
be ubiquitous in dwarfs \cite{Grebel2004}.  In low-mass galaxies the
younger and/or metal-rich populations tend to be more centrally
concentrated (e.g., \cite{Grebel1997, Hurley-Keller1999, Zaritsky2000,
Harbeck2001, Crnojevic2010, Lianou2010, Lisker2006a, Kniazev2009,
Crnojevic2011a, Kirby2011a, Weisz2011}).  Similarly, more metal-rich,
younger populations tend to be dynamically colder than older, more
metal-poor populations (e.g., \cite{Tolstoy2004, Battaglia2006,
Ibata2006, Battaglia2011}), though not all dwarf galaxies show such
population gradients (e.g., \cite{Koch2007a, Koch2007b, Ural2010}). 

Taking advantage of the wealth of information available for nearby,
low-metallicity dwarf galaxies and their identifiable, metal-poor old
populations, they will be in the focus of this review (although we
will also consider galaxies beyond the Local Group).  Many of the
Local Group dwarf galaxies are even close enough for detailed
spectroscopic abundance analyses of individual member stars, revealing
not only the overall metal content, but the star-to-star variations in
light and heavy elements of different nucleosynthetic origin (e.g.,
\cite{Aden2011, Aoki2009, Cohen2009, Cohen2010, deBoer2012, Hill2000,
Kalirai2009, Kirby2011a, Kirby2011b, Koch2006, Koch2007c, Koch2008a,
Koch2008b, Norris2008, Sadakane2004, Shetrone2001, Shetrone2003}),
which in turn allow us to draw inferences about the conditions under
which these metal-deficient galaxies and their stars formed.

\section{2. Metal-Poor Galaxies Along the Metallicity-Luminosity Relation}

As pointed out by \cite{Tassis2012}, star formation and molecular gas
are closely related, and the ability to form molecular gas depends on
metallicity (including the presence of dust to provide shielding).
These authors argue that in the absence of metals, molecular gas
cannot easily form, leading to a long gas consumption time scale, a
low star formation efficiency, and slow enrichment.  Above a critical
metallicity threshold, however, both star formation rate and
enrichment accelerate, resulting in a rapid increase in stellar mass
and metallicity \cite{Tassis2012}.  Thus, without significant
pre-enrichment by Population III stars, one would \emph{not} expect to
observe a continuously occupied metallicity-luminosity relation, but
instead to see a bimodality in metallicity \cite{Tassis2012}.   

Instead, the metallicity-luminosity relation for dwarf galaxies
smoothly extends the correlation found for giant galaxies all the way
down to ultra-faint dwarf spheroidal galaxies with luminosities as low
as $M_V \sim -3$ mag \cite{McConnachie2012}.  Particularly for dwarf
irregular and dwarf spheroidal galaxies, there is also an increasing
``mean age'' trend in the sense that the more metal-poor dwarf
galaxies tend to be dominated by stellar populations with older ages
(see also \cite{Paudel2010}; in contrast to the behavior seen in,
e.g., giant elliptical galaxies).  At optical wavelengths, the
vigorously star-forming dwarf galaxies discussed in the next paragraph
differ from this trend.  Generally, the scatter in the
metallicity-luminosity relation increases with decreasing luminosity
(\cite{McConnachie2012}; see also \cite{Panter2008}).

\subsection{2.1 Blue Compact Dwarf Galaxies}

Metal-deficient star-forming dwarf galaxies, in particular blue
compact dwarfs (BCDs), appear to be too luminous for their low
metallicities \cite{Kunth2000, Kniazev2003}, thus deviating from the
mean locus of dwarf galaxies in the metallicity-luminosity relation.
Their apparent overluminosity is likely largely an effect of the
pronounced starbursts they are undergoing at the present time.  The
analysis of the light and heavy element content of H\,{\sc ii} regions
in such metal-poor BCDs and the low dispersion of the resulting
element abundance ratios suggests a probable primary origin of these
elements in massive stars \cite{Izotov1999}. This used to be
considered an argument in favor of these galaxies being ``young'' in
the sense that they are now undergoing their first burst of star
formation (e.g., \cite{Izotov1999}).  

The most metal-deficient BCD known, I~Zw~18, which has a metallicity
of only $Z \sim 1/50$, has often been considered an epitome of a truly
young galaxy in the present-day Universe.  \emph{Hubble Space
Telescope} (HST) photometry of resolved stars in I~Zw~18 seemed to
support the ``young galaxy'' scenario, since those data showed no
trace of an older red giant branch (RGB) population \cite{Izotov2004}.
Even deeper HST photometry, however, revealed that I~Zw~18 is more
distant than previously assumed and contains a well-defined RGB,
demonstrating that this extremely metal-poor galaxy is not primordial
\cite{Aloisi2007}.   

More generally, recent studies of the integrated light have confirmed
that BCDs are old systems currently experiencing starbursts (e.g.,
\cite{Zhao2011}).  The stellar mass-weighted ages were found to be as
old as up to 10 Gyr, while the luminosity-weighted ages are of the
order of only 10 Myr \cite{Zhao2011}.  The current star formation
rates exceed the averaged past star formation rates by factors of more
than two to three \cite{Zhao2011}.  The colors of BCDs are best
reproduced when invoking continuous star formation and a recent burst
\cite{Zitrin2009}, although theoretical chemical evolution models also
suggest series of bursts in combination with metal-enhanced winds
\cite{Yin2011}.  Observational data suggest that the strongly enhanced
star formation activity of BCDs is often caused by mergers or
interactions with a companion, whereas one possible mechanism to halt
the burst activity are supernova-driven winds \cite{Perez2011}.
Theoretical models support gas-rich dwarf-dwarf mergers, whose old
populations turn into diffuse low-surface-brightness components
\cite{Bekki2008}.  The starbursts in the compact cores of BCDs may be
fueled by low-metallicity gas from the outer extended gas disks of the
progenitor dwarfs, resulting in metal-poor young stars
\cite{Bekki2008}.    

\subsection{2.2 Tidal Dwarf Galaxies}

Tidal dwarf galaxies are long-lived, gravitationally bound objects
free of dark matter that form from ``recycled'', mainly gaseous
material previously torn out during interactions between massive
(spiral) galaxies (see \cite{Bournaud2010} and references therein).
The most massive tidal dwarfs are believed to form from matter
concentrations at the tips of tidal tails and will experience
starburst activity comparable to that of BCDs \cite{Bournaud2010}.

Tidal dwarf galaxies also deviate from the metallicity-luminosity
relation.  At a given luminosity, their metallicity {\em exceeds} that
of regular dwarf galaxies since they are metal-rich already ``at
birth'', forming from the pre-enriched gas of much more massive
galaxies \cite{Duc2000}.  In fact, their average metallicity (1/3
solar) is essentially independent of their absolute luminosity
\cite{Duc2000}.  The star formation efficiency in tidal dwarfs
resembles that of spiral galaxies \cite{Braine2001}.  This may be
attributed to the improved self-shielding due to their
higher-than-normal metallicities (as compared to other dwarf
galaxies), facilitating the formation of molecular gas and star
formation \cite{Braine2001}.  HST imaging suggests that the tidal
dwarf candidates in the M81 group contain young stars formed in situ,
but also older RGB stars that may have formed in the massive galaxies
of this group \cite{Makarova2002, deMello2008}.  

While too high a metallicity for a given luminosity is a good
indicator for young tidal dwarf candidates,  it is more difficult to
identify {\em old} tidal dwarfs.  Depending on when they formed, these
objects may be as metal-poor as ``normal'' dwarf irregulars
\cite{Hunter2000}.

\subsection{2.3 Dwarf Irregular and Dwarf Elliptical/Spheroidal Galaxies}

In contrast to the previously discussed dwarfs (Sections 2.1 and 2.2),
dwarf irregular (dIrr), dwarf elliptical (dE), and dwarf spheroidal
(dSph) galaxies typically follow a well-defined metallicity-luminosity
relation (see, e.g., \cite{McConnachie2012}).  DSph galaxies usually
host prominent old populations \cite{Grebel2004}, and the mean
metallicities of these systems are thus dominated by the metallicities
of their (old and intermediate-age) RGB populations.  DIrrs, on the
other hand, have typically experienced continuous star formation with
some amplitude fluctuations \cite{Tosi1991}.  This mode of extended
periods of star formation interrupted by quiescent phases is also
called ``gasping'' star formation \cite{Cignoni2010}.  That the
(predominantly old) dSphs follow the same global
metallicity-luminosity (or metallicity-stellar mass) relation as the
dIrrs suggests that the dSphs did not experience substantial tidal
stripping despite their proximity to massive galaxies
\cite{Gilmore2012}.  Moreover, with the exception of dSphs that are
currently being disrupted, their structure and kinematics do not show
evidence of their being unbound tidal remnants (e.g.,
\cite{Odenkirchen2001, Klessen2003, Gilmore2007, Donato2009,
Walker2011}). 

A closer investigation of the metallicity-luminosity relation reveals
that there is an offset between dSphs and dIrrs (e.g.,
\cite{Richer1998}), which persists even when limiting the comparison
to the metallicities of the old populations in the two types of dwarfs
\cite{Grebel2003}.  Despite their older mean ages, dSph galaxies are
more metal-rich than expected from their luminosities, which may imply
that they experienced initially more vigorous star formation and
enrichment than the slowly evolving dIrrs \cite{Grebel2003}.  At a
given stellar mass, the metallicities of dSphs are typically higher by
a factor of three than those of dIrrs \cite{Woo2008}.  This makes it
difficult for dIrrs to evolve into dSphs if their star formation were
to cease since mere passive fading would require more than a Hubble
time, though such an evolution appears to be possible for
low-luminosity transition-type dIrr/dSph galaxies \cite{Grebel2003}.
DSphs and dIrrs may thus be intrinsically different and formed under
different conditions \cite{Grebel2003}.  Alternatively, dSphs may have
lost a large amount of their initial mass due to later tidal (e.g.,
\cite{Kravtsov2004}) and/or ram pressure stripping (e.g.,
\cite{Grebel2003}) while allowing the galaxies to survive as
dark-matter-dominated bound entities (e.g., \cite{Wolf2010} and Sect.\
3).  

The low-luminosity end of the metallicity-luminosity relation consists
exclusively of the recently discovered faint and ultra-faint ($M_V <
-8$ mag) dSph galaxies around the Milky Way and M31 (e.g.,
\cite{Aden2009a, Belokurov2006, Harbeck2005, McConnachie2008,
Martin2006, Richardson2011, Walsh2007, Zucker2004, Zucker2006a,
Zucker2006b, Zucker2007}).  The ultra-faint dSphs are not only of
interest for the cosmological substructure or missing satellite
problem (e.g., \cite{Kravtsov2010, Madau2008, Tollerud2008}), but are
also intriguing objects from the point of view of galaxy evolution:
They are the least massive, least luminous, most metal-poor,
dark-matter-dominated galaxies known.  As objects containing only old,
metal-deficient stars, they may be potential ``pre-re-ionization
fossils'' \cite{Aden2009b, Bovill2009, Gnedin2006, Madau2008,
Ricotti2010, Okamoto2012}.

\section{3. Re-Ionization and Later Environmental Effects}

Environment, i.e., the local galaxy density, the degree of proximity
to massive galaxies, the degree of activity in massive galaxies, and
immediate galaxy interactions, affect the evolution and properties of
low-mass galaxies.  The earlier mentioned morphology-density relation
is a prominent example of the impact of environment.  One of the
intriguing questions in this context, particularly with respect to
dwarf galaxies, is the importance of a galaxy's intrinsic properties
(``nature'') vs.\ the importance of external influences (``nurture'').
Clearly, both play a role in shaping dwarf galaxies in groups and
clusters (e.g., \cite{Annibali2011, Conselice2003, Crnojevic2011b,
Crnojevic2012, DeRijcke2010, Girardi2003, Lisker2006b, Lisker2007,
Karachentsev2002a, Karachentsev2002b, Karachentsev2003a,
Karachentsev2003b, Karachentsev2003c, Pasetto2011, Paudel2010,
Zandivarez2011}).  

In the Local Group and other nearby groups, early-type companions with
H\,{\sc i} masses below $10^6$~M$_{\odot}$ are usually found within
$\sim 300$ kpc around massive galaxies, whereas late-type dwarfs with
$M_{\rm HI} > 10^7$~M$_{\odot}$ are mostly located at larger distances
\cite{Grebel2003, Grcevich2009}.  Low-mass transition-type dIrr/dSph
galaxies fill the range in between these properties \cite{Grebel2003}.
This seems to suggest that gas removal caused by various processes
related to the proximity to massive galaxies may be the cause for the
morphological segregation (see also \cite{vandenBergh1994,
vandenBergh1999}), in particular ram-pressure stripping, although the
present-day Galactic halo gas densities appear to be too low
\cite{Grebel2003}.  

Cosmological simulations taking into account gas cooling, star
formation, supernova (SN) feedback, enrichment, and ultraviolet
heating lead to satellite galaxies of which 95\% are gas-free at the
present time \cite{Sawala2012}.  According to these simulations,
satellites need total masses of at least $5\cdot10^9$~M$_{\odot}$ in
order to retain their gas.  Interestingly, also isolated dwarf
galaxies are predicted to be largely gas-deficient at the present time
if their total masses are below $\sim 10^9$~M$_{\odot}$
\cite{Sawala2012}.  These authors conclude that while gas stripping
aids in removing gas from dwarf satellites, the total mass of a dwarf
galaxy (thus an intrinsic property) is the primary factor determining
whether a galaxy retains or loses its gas, and the total mass also
governs a galaxy's star formation and enrichment.  

In contrast, other models advocate a stronger role of galaxian
environment.  For instance, in simplified models presented by
\cite{Nichols2011},  feedback-assisted ram pressure stripping or tidal
stripping during the period when infalling dwarfs still experienced
active star formation may have effectively removed the gas,
reproducing the observed morphological segregation.  \cite{Ocvirk2011}
investigate the effects of re-ionization on satellite properties.  They
find that ``external'', uniform re-ionization from the cosmic radiation
field may have less of an effect than ``internal'', distance-dependent
re-ionization from the most massive Milky Way progenitor.  Assuming
that the satellites were already at similar distances as today, the
photoevaporation driven by the Galactic radiation field first removed
star-forming material from the inner satellites, allowing the outer
ones to continue star formation for a longer period
(\cite{Ocvirk2011}; see also \cite{vandenBergh1994}).  This
``internal'' scenario can reproduce the observed cumulative radial
satellite  distribution very well, in contrast to a scenario that only
considers the external ionization field \cite{Ocvirk2011}. 

\cite{Munoz2009} include four epochs of star formation in their
simulations:  (1) At $z \sim 20$ star formation occurs in systems
large enough for H$_2$ cooling resulting in very metal-poor stars, (2)
later through H\,{\sc i} cooling (leading to re-ionization), (3) during
re-ionization until $z \sim 2$ through further H\,{\sc i} cooling in
subhalos that are large enough not yet to have lost their gas, and
finally (4) during the last 10 Gyr through metal cooling.  Ultra-faint
dSphs with $M_V > -5$ then formed very early in low-mass halos via H$_2$
cooling \cite{Munoz2009}.  The H$_2$ cooling threshold of $M_{\rm H2}
\sim 10^5$ to $10^6$~M$_{\odot}$ implies a luminosity threshold, which
excludes the existence of ever fainter satellites.  Satellites in the
luminosity range of $-5 > M_V > -9$ experienced star formation via
H\,{\sc i} cooling and photo-heating feedback, whereas most of the
stars in the brighter satellites formed after re-ionization
\cite{Munoz2009}.

\section{4. Clues from Chemical Abundances}

Dwarf galaxies do not represent a simple stellar population of a
single age and metallicity.  Instead, they usually experience extended
episodes of star formation leading to gradual enrichment and,
depending on duration of star formation, also to a measurable range of
ages (e.g., \cite{Grebel1999, Grebel2004, Ikuta2002, Marcolini2008,
Pasetto2012}).  For nearby dwarfs the spread in metallicity (and even
in individual element abundance ratios) can be measured directly
through absorption line spectroscopy of individual stars, particularly
along the RGB.  In more distant dwarfs, the brightest supergiants may
still be accessible (e.g., \cite{Bresolin2007, Tautvaisiene2007,
Venn2001}), while emission-line spectroscopy of planetary nebulae or
H\,{\sc ii} regions can be conducted over a wide range of distances
(e.g., \cite{Kniazev2004, Kniazev2005, Kniazev2007, Kniazev2008,
Lee2003, Lee2007, Magrini2005a, Magrini2005b, Pilyugin2001,
Pilyugin2012, vanZee2006}).  

\subsection{4.1 Metallicity Distribution Functions and Gradients}

Recent spectroscopic analyses of large numbers of RGB stars in nearby
dwarf galaxies resulted in well-sampled metallicity distribution
functions (MDFs) for many dSphs and dIrrs.  Often these MDFs show a
gradual rise toward higher metallicities and then a steeper fall-off,
but exceptions exist (e.g., \cite{Battaglia2006, Battaglia2011,
Bosler2007, Carrera2008a, Carrera2008b, Chou2007, Kirby2011a,
Koch2006, Koch2007b, Koch2007c, Leaman2009, Norris2008}).  A G-dwarf
problem (or, more accurately, a K-giant problem) is found in all
dwarfs (e.g., \cite{Koch2007c}).  A comparison of the MDFs with simple
chemical evolution models generally shows a poor fit for closed-box
models or pristine gas models, but better results for pre-enrichment,
leaky box, or models with additional gas infall (e.g., \cite{Koch2006,
Kirby2011a}).  The range in stellar metallicity increases with
galaxian luminosity (or stellar mass) \cite{Kirby2011a}.  The
interplay between slow gas infall, low-efficiency star formation, and
strong galactic winds can reproduce both the observed MDFs and element
abundance ratios \cite{Lanfranchi2010}.  The pronounced decrease of
the metal-rich end of the MDF is then caused by the removal of gas --
without winds, the resulting MDF would be much more metal-rich than
observed \cite{Lanfranchi2010, Kirby2011b}.  Other models, however, do
not require strong winds but employ external mechanisms for the loss
of heated, enriched gas such as ram pressure or tidal interactions
\cite{Marcolini2006}. 

Shallow gradients of decreasing metallicity with galactocentric radius
are commonly found in spectroscopic surveys \cite{Kirby2011a},
confirming the finding that more metal-rich and/or younger populations
are more centrally concentrated and kinematically colder (see Section
1 and, e.g., \cite{Harbeck2001}), although there are also exceptions
(e.g., \cite{Koch2007b, Koch2007c, Ural2010}).  The lack of a gradient
may be due to insufficient radial area coverage in spectroscopic
surveys \cite{Kirby2011a} or be caused by stripping (see
\cite{Sales2010}).  Tidal stripping would primarily remove the more
extended (and more metal-poor) stellar populations and homogenize the
stellar radial velocities of the components \cite{Sales2010}.
\cite{Ural2010}, on the other hand, point out the danger of
overinterpreting small stellar samples in favor of the existence of a
gradient.

A number of theoretical models have explored the possible origin of
population gradients.  For example, apart from SN-driven blow-outs SNe
in the outer regions of dwarfs may drive inward-propagating winds that
ultimately concentrate cold gas in the inner regions \cite{Mori2002}.
Star formation may re-start from this enriched gas \cite{Mori2002},
which would result in a metallicity gradient.  Other models suggest
that any gradients will be quickly erased since the winds of SNe II
tend to homogenize the interstellar medium \cite{Marcolini2006,
Marcolini2008}.  Metallicity gradients are then introduced by the
spatially inhomogeneous enrichment through the much rarer SNe Ia,
which occur more frequently in the denser populated central regions
\cite{Marcolini2008}.  \cite{Pasetto2010} show that the time scales of
the change of the dark matter profiles of dwarf galaxies (from cuspy
to cored profiles) due to feedback and the redistribution of
(baryonic) matter on the one hand and of the formation of chemical
gradients on the other hand may be linked.  The oldest stars in the
dwarf move on radial orbits, and inner and outer regions are thus well
mixed (similar to \cite{Marcolini2008}, but due to a different
mechanism).  The evolving density profile implies that younger, more
metal-rich stars forming in the central regions tend to be on more
circular orbits with smaller radial velocity dispersion, resulting in
less mixing with the outskirts \cite{Pasetto2010}. 

\subsection{4.2 Element Abundance Trends and Inhomogeneities}

Dwarf galaxies exhibit considerable stellar abundance spreads.  The
range of metallicities often spans $> 1$~dex in [Fe/H] even in
galaxies dominated by old populations (e.g., \cite{Shetrone2001,
Geisler2005, Grebel2003, Norris2008, Cohen2010, Lai2011, Kirby2011a}.
In several dwarf galaxies studied in detail it turns out that at {\em
any} given age there is a spread in metallicity (e.g.,
\cite{Glatt2008b, Haschke2012a, Haschke2012b, Kayser2009, Kniazev2005,
Koch2007c, Norris2008}).  Moreover, at a given overall metallicity one
tends to find scatter in the $\alpha$ abundance ratios (e.g.,
\cite{Koch2008a, Koch2008b, Aden2011}).  These galaxies were not
well-mixed when these stars formed.  The trends as such differ from
galaxy to galaxy, confirming once again that no two dwarfs are alike
-- not even dwarfs of the same luminosity and morphological type
\cite{Grebel1997}. 

There are a number of theoretical models addressing these issues.  It
seems that at very early times and very low metallicities star
formation was governed by stochasticity and inhomogeneous
heavy-element pollution caused by the few early SNe II
\cite{Marcolini2008}.  Subsequently, with increasing enrichment, the
feedback from these SNe led to fairly homogeneous mixing
\cite{Marcolini2008}.  At higher metallicities, individual SNe Ia
again led to localized inhomogeneous abundance patterns
\cite{Marcolini2008}, resulting in a substantial element abundance
spread even for coeval stars, just as seen in the observations (see
also \cite{Pasetto2010}).     

Considering trends in [$\alpha$/Fe] vs.\ [Fe/H], stars in dwarf
galaxies resemble those in the Galactic halo at low metallicity
([Fe/H]~$< -2$ dex) (e.g., \cite{Cohen2010, Koch2008a, Lai2011,
Venn2012}), but tend to show lower [$\alpha$/Fe] ratios at higher
metallicities as compared to typical Galactic halo stars (e.g.,
\cite{Shetrone2001}).  SNe Ia thus seem to contribute already at lower
metallicities in dwarfs than in our halo.  See \cite{deBoer2012} for a
nice demonstration of the age dependence of the [$\alpha$/Fe] vs.\
[Fe/H] trend.  The different locations of the ``turn-over''
from the near-constant [$\alpha$/Fe] towards lower ratios appears to
roughly depend on a galaxy's stellar mass.  

Lower (than Galactic) [$\alpha$/Fe] ratios at a given [Fe/H] may
indicate (1) low star formation rates (with few contributions of
$\alpha$ elements from SNe II), (2) a substantial loss of SN ejecta
(metals) via galactic winds, or (3) a larger contribution from SNe Ia
(enhancing the Fe content with respect to $\alpha$ elements)
\cite{Shetrone2001}.  Several studies show evidence for early rapid
enrichment based on heavy neutron-capture and r-process element
abundance ratios (e.g., \cite{Aden2011}), while low-efficiency,
extended star formation with stochastic, inhomogeneous contributions
from SNe Ia ejecta and asymptotic giant branch stars appears to have
been important at higher metallicities (e.g., \cite{Aden2011,
Cohen2010, Geisler2005, Venn2012}).  Also, an initial mass function
sparsely populated with massive stars due to the low star formation
rate may contribute to the slow build-up of metallicity and to the
inhomogeneities \cite{Koch2008a}.  

\subsection{4.3 Extremely Metal-Poor Stars}

Major efforts were devoted to the search for extremely metal-poor
stars in our Milky Way and in nearby dwarf galaxies.  While no stars
as metal-deficient as found in the Galactic halo (\cite{Caffau2012}
and references therein) have been detected in Galactic satellites yet,
the current record holders have [Fe/H]~$= -3.96\pm0.06$ dex in the
dSph Sculptor \cite{Tafelmeyer2010} and $-2.67 \pm 0.33$ dex in the
irregular Large Magellanic Cloud (LMC) \cite{Haschke2012b}.   

Ultra-faint dSphs appear to be a particularly rewarding hunting ground
for very metal-deficient stars as suggested by both observational data
(e.g., \cite{Lai2011, Norris2008, Norris2010, Simon2010}) and
simulations (e.g., \cite{Munoz2009, Salvadori2009}).  \cite{Simon2010}
and \cite{Koch2008b} argue that the abundance patterns in extremely
metal-poor stars in the ultra-faint dSphs Leo\,IV and Hercules are
consistent with enrichment through Population III SN explosions.
According to simulations, ultra-faint dSphs may have experienced star
formation via H$_2$ cooling of pristine gas at high redshift
(\cite{Munoz2009}; Sect.\ 3), an inefficient process resulting in low
star formation rates \cite{Salvadori2009}.  These strongly
dark-matter-dominated galaxies themselves may be merger products of
very early H$_2$-cooling minihalos \cite{Salvadori2009}, which may
account for their metallicity spread.

While the element abundance ratios in dSphs were once considered an
argument against a significant contribution of such galaxies to the
build-up of the Galactic halo (see, e.g., \cite{Geisler2007} and
references therein), this picture has changed recently.  A number
of studies point out the good agreement of the abundance ratios in
metal-poor stars in the Galactic halo and in dSphs.  This chemical
similarity makes contributions from dSphs plausible and indicates that
the early chemical evolution of galaxies on all scales may be similar
(e.g., \cite{Cohen2010, Frebel2010, Lai2011, Norris2010, Simon2010}).

Some models predict that most of the stars in the stellar halo
(particularly the inner halo) were formed not in dSphs, but in a small
number of massive dIrr-like galaxies that then were accreted by the
Milky Way some 10 Gyr ago \cite{Robertson2005}.  The element abundance
ratios in genuinely old, metal-poor field and globular cluster stars
in the Magellanic Clouds turn out to be in good agreement with
Galactic halo abundances \cite{Johnson2006, Mucciarelli2010,
Haschke2012b}.  These findings are compatible with the predicted
early accretion of dIrrs.   

Ultra-faint dSphs, on the other hand, may be a very important source
of extremely metal-deficient stars in the Galactic halo.
Intriguingly, and contrary to earlier views, these galaxies may even
contain {\em too many} such extremely metal-poor stars as compared to
the Galactic halo \cite{Lai2011, Starkenburg2010}.  However, as
emphasized by \cite{Lai2011}, most halo star studies focused on the
\emph{inner} halo.  Carbon-enhanced metal-poor (CEMP) stars may serve
as a tracer for generally very metal-poor stars, and their number
increases with decreasing metallicity and with height above the
Galactic plane \cite{Carollo2012}. In the {\em outer} halo the
fraction of CEMP stars is approximately twice as much as in the inner
halo \cite{Carollo2012}.  Hence we may expect also a larger (yet to be
detected) number of extremely metal-poor stars there.  

Combining these findings and arguments, a picture emerges where the
inner halo may consist largely of stars contributed early on by larger
progenitor systems (which make it difficult to uncover dSph
contributions), while the outer halo may plausibly contain a
significant stellar component accreted from small dwarfs such as dSphs
(\cite{Lai2011}; see also \cite{Bell2008}).  Clearly, detailed element
abundance information for more old stars in the outer halo, in the
Magellanic Clouds, and in ultra-faint and classical dSphs is desirable
to explore the importance of early accretion and of the origin of very
metal-poor stars.



\end{document}